\documentclass[12pt]{article}

\usepackage[body={17.5cm, 21cm},right=2cm]{geometry}

\usepackage{color}
\usepackage{graphicx}
\usepackage{epsf}
\usepackage{graphicx,epsfig}

\usepackage{multirow}

\usepackage{chemarrow}
\usepackage[intlimits]{amsmath}

\usepackage{amsmath}
\usepackage{amssymb}
\usepackage{epsfig}
\usepackage{cite}
\usepackage{color,colordvi}

\newcommand{\bi}{\begin{itemize}}
\newcommand{\ei}{\end{itemize}}

\newcommand{\bx}{{\bf{x}}}

\newcounter{hran}

\def\MSbar{\relax\ifmmode\overline{\rm MS}\else{$\overline{\rm MS}${ }}\fi}

\def\d{\rm d}

\def\p{\partial}

\def\tr{{\rm tr}}

\numberwithin{equation}{section}
\begin{document}
\vspace{5mm}
\vspace{0.5cm}
\begin{center}

\def\thefootnote{\fnsymbol{footnote}}

\rightline{ ICCUB-12-366}
\bigskip
\bigskip
\bigskip

{\large \bf 
Global Supersymmetry on Curved Spaces \\
in Various Dimensions}
\\[1.5cm]
{\large  Alex Kehagias$^{1}$ and Jorge G. Russo$^{2,3}$}
\\[0.5cm]

\vspace{.3cm}
{\normalsize {\it  $^{1}$ Physics Division, National Technical University of Athens, \\15780 Zografou Campus, Athens, Greece.}}\\

\vspace{.3cm}
{\normalsize { \it $^{2}$  Instituci\'o Catalana de Recerca i Estudis Avan\c cats (ICREA), \\
Pg. Lluis Companys, 23, 08010 Barcelona, Spain.\\
$^{3}$
 Institute of Cosmos Sciences and Estructura i Constituents de la Materia, \\
Facultat de F\'\i sica, Universitat de Barcelona, \\
Marti i Franqu\`es, 1, 08028 Barcelona, Spain. 
 }}\\

\vspace{.3cm}


\end{center}

\vspace{2.5cm}

\hrule \vspace{0.3cm}
{\small  \noindent \textbf{Abstract} \\[0.3cm]
\noindent 
We propose  methods towards a systematic determination  of
$d$ dimensional curved spaces  where  Euclidean field theories with rigid 
supersymmetry can be defined. 
The analysis is carried out from a group theory as well as from a supergravity
point of view. In particular, by using  appropriate gauged supergravities  in various dimensions we show  that 
supersymmetry 
can  be defined in conformally flat spaces, such as non-compact hyperboloids 
$\mathbb{H}^{n+1}$ and compact spheres $\mathbb{S}^n$ or --by turning on appropriate Wilson lines corresponding 
to R-symmetry vector fields-- on $\mathbb{S}^1\times \mathbb{S}^{n}$, with $n<6$. 
By group theory arguments we show that  
Euclidean field theories with rigid supersymmetry cannot be consistently defined on
round spheres $\mathbb{S}^d$ 
if $d>5$ (despite the existence of Killing spinors). 
We also show that distorted spheres and certain orbifolds are also allowed by the group theory classification. 

\vspace{0.5cm}  \hrule
\vskip 1cm

\def\thefootnote{\arabic{footnote}}
\setcounter{footnote}{0}



\baselineskip= 19pt

\newpage 




\newcommand{\fix}{\Phi(\mathbf{x})}
\newcommand{\fiLx}{\Phi_{\rm L}(\mathbf{x})}
\newcommand{\fiNLx}{\Phi_{\rm NL}(\mathbf{x})}
\newcommand{\fik}{\Phi(\mathbf{k})}
\newcommand{\fiLk}{\Phi_{\rm L}(\mathbf{k})}
\newcommand{\fiLkone}{\Phi_{\rm L}(\mathbf{k_1})}
\newcommand{\fiLktwo}{\Phi_{\rm L}(\mathbf{k_2})}
\newcommand{\fiLkthree}{\Phi_{\rm L}(\mathbf{k_3})}
\newcommand{\fiLkfour}{\Phi_{\rm L}(\mathbf{k_4})}
\newcommand{\fiNLk}{\Phi_{\rm NL}(\mathbf{k})}
\newcommand{\fiNLkone}{\Phi_{\rm NL}(\mathbf{k_1})}
\newcommand{\fiNLktwo}{\Phi_{\rm NL}(\mathbf{k_2})}
\newcommand{\fiNLkthree}{\Phi_{\rm NL}(\mathbf{k_3})}

\newcommand{\kernel}{f_{\rm NL} (\mathbf{k_1},\mathbf{k_2},\mathbf{k_3})}
\newcommand{\dirac}{\delta^{(3)}\,(\mathbf{k_1+k_2-k})}
\newcommand{\dirackonektwokthree}{\delta^{(3)}\,(\mathbf{k_1+k_2+k_3})}

\newcommand{\beq}{\begin{equation}}
\newcommand{\eeq}{\end{equation}}

\newcommand{\be}{\begin{equation}}
\newcommand{\ee}{\end{equation}}

\newcommand{\bea}{\begin{eqnarray}}
\newcommand{\eea}{\end{eqnarray}}

\newcommand{\angk}{\hat{k}}
\newcommand{\angn}{\hat{n}}

\newcommand{\tfnow}{\Delta_\ell(k,\eta_0)}
\newcommand{\tf}{\Delta_\ell(k)}
\newcommand{\tfone}{\Delta_{\el\ell_1}(k_1)}
\newcommand{\tftwo}{\Delta_{\el\ell_2}(k_2)}
\newcommand{\tfthree}{\Delta_{\el\ell_3}(k_3)}
\newcommand{\tffour}{\Delta_{\el\ell_1^\prime}(k)}
\newcommand{\deltatilde}{\widetilde{\Delta}_{\el\ell_3}(k_3)}

\newcommand{\alm}{a_{\ell m}}
\newcommand{\almL}{a_{\ell m}^{\rm L}}
\newcommand{\almNL}{a_{\ell m}^{\rm NL}}
\newcommand{\almone}{a_{\el\ell_1 m_1}}
\newcommand{\almLone}{a_{\el\ell_1 m_1}^{\rm L}}
\newcommand{\almNLone}{a_{\el\ell_1 m_1}^{\rm NL}}
\newcommand{\almtwo}{a_{\el\ell_2 m_2}}
\newcommand{\almLtwo}{a_{\el\ell_2 m_2}^{\rm L}}
\newcommand{\almNLtwo}{a_{\el\ell_2 m_2}^{\rm NL}}
\newcommand{\almthree}{a_{\el\ell_3 m_3}}
\newcommand{\almLthree}{a_{\el\ell_3 m_3}^{\rm L}}
\newcommand{\almNLthree}{a_{\el\ell_3 m_3}^{\rm NL}}

\newcommand{\YLMstar}{Y_{L M}^*}
\newcommand{\Ylmstar}{Y_{\ell m}^*}
\newcommand{\Ylmstarone}{Y_{\el\ell_1 m_1}^*}
\newcommand{\Ylmstartwo}{Y_{\el\ell_2 m_2}^*}
\newcommand{\Ylmstarthree}{Y_{\el\ell_3 m_3}^*}
\newcommand{\Ylmstarfour}{Y_{\el\ell_1^\prime m_1^\prime}^*}
\newcommand{\Ylmstarfive}{Y_{\el\ell_2^\prime m_2^\prime}^*}
\newcommand{\Ylmstarsix}{Y_{\el\ell_3^\prime m_3^\prime}^*}

\newcommand{\YLM}{Y_{L M}}
\newcommand{\Ylm}{Y_{\ell m}}
\newcommand{\Ylmone}{Y_{\el\ell_1 m_1}}
\newcommand{\Ylmtwo}{Y_{\el\ell_2 m_2}}
\newcommand{\Ylmthree}{Y_{\el\ell_3 m_3}}
\newcommand{\Ylmfour}{Y_{\el\ell_1^\prime m_1^\prime}}
\newcommand{\Ylmfive}{Y_{\el\ell_2^\prime m_2^\prime}}
\newcommand{\Ylmsix}{Y_{\el\ell_3^\prime m_3^\prime}}

\newcommand{\comm}[1]{\textbf{\textcolor{rossos}{#1}}}
\newcommand{\lsim}{\,\raisebox{-.1ex}{$_{\textstyle <}\atop^{\textstyle\sim}$}\,}
\newcommand{\gsim}{\,\raisebox{-.3ex}{$_{\textstyle >}\atop^{\textstyle\sim}$}\,}

\newcommand{\jl}{j_\ell(k r)}
\newcommand{\jlfourone}{j_{\el\ell_1^\prime}(k_1 r)}
\newcommand{\jlfivetwo}{j_{\el\ell_2^\prime}(k_2 r)}
\newcommand{\jlsixthree}{j_{\el\ell_3^\prime}(k_3 r)}
\newcommand{\jlsix}{j_{\el\ell_3^\prime}(k r)}
\newcommand{\jlthree}{j_{\el\ell_3}(k_3 r)}
\newcommand{\jlthreetau}{j_{\el\ell_3}(k r)}

\newcommand{\Gaunt}{\mathcal{G}_{\el\ell_1^\prime \, \el\ell_2^\prime \, 
\el\ell_3^\prime}^{m_1^\prime m_2^\prime m_3^\prime}}
\newcommand{\Gaunttwo}{\mathcal{G}_{\el\ell_1^\prime \, \el\ell_2^\prime \, 
\el\ell_3}^{m_1^\prime m_2^\prime m_3}}
\newcommand{\Gauntstardef}{\mathcal{H}_{\el\ell_1 \, \el\ell_2 \, \el\ell_3}^{m_1 m_2 m_3}}
\newcommand{\Gauntstarone}{\mathcal{G}_{\el\ell_1 \, L \,\, \el\ell_1^\prime}
^{-m_1 M m_1^\prime}}
\newcommand{\Gauntstartwo}{\mathcal{G}_{\el\ell_2^\prime \, \el\ell_2 \, L}
^{-m_2^\prime m_2 M}}

\newcommand{\de}{{\rm d}}

\newcommand{\dangn}{d \angn}
\newcommand{\dangk}{d \angk}
\newcommand{\dangkone}{d \angk_1}
\newcommand{\dangktwo}{d \angk_2}
\newcommand{\dangkthree}{d \angk_3}
\newcommand{\dk}{d^3 k}
\newcommand{\dkone}{d^3 k_1}
\newcommand{\dktwo}{d^3 k_2}
\newcommand{\dkthree}{d^3 k_3}
\newcommand{\dkfour}{d^3 k_4}
\newcommand{\dallk}{\dkone \dktwo \dk}

\newcommand{\FT}{ \int  \! \frac{d^3k}{(2\pi)^3} 
e^{i\mathbf{k} \cdot \angn \eta_0}}
\newcommand{\planewave}{e^{i\mathbf{k \cdot x}}}
\newcommand{\dallkfourier}{\frac{\dkone}{(2\pi)^3}\frac{\dktwo}{(2\pi)^3}
\frac{\dkthree}{(2\pi)^3}}

\newcommand{\Bis}{B_{\el\ell_1 \el\ell_2 \el\ell_3}^{m_1 m_2 m_3}}
\newcommand{\Avbis}{B_{\el\ell_1 \el\ell_2 \el\ell_3}}

\newcommand{\los}{\mathcal{L}_{\el\ell_3 \el\ell_1 \el\ell_2}^{L \, 
\el\ell_1^\prime \el\ell_2^\prime}(r)}
\newcommand{\loszero}{\mathcal{L}_{\el\ell_3 \el\ell_1 \el\ell_2}^{0 \, 
\el\ell_1^\prime \el\ell_2^\prime}(r)}
\newcommand{\losone}{\mathcal{L}_{\el\ell_3 \el\ell_1 \el\ell_2}^{1 \, 
\el\ell_1^\prime \el\ell_2^\prime}(r)}
\newcommand{\lostwo}{\mathcal{L}_{\el\ell_3 \el\ell_1 \el\ell_2}^{2 \, 
\el\ell_1^\prime \el\ell_2^\prime}(r)}
\newcommand{\losfNL}{\mathcal{L}_{\el\ell_3 \el\ell_1 \el\ell_2}^{0 \, 
\el\ell_1 \el\ell_2}(r)}



\def\d{d}
\def\C{{\rm CDM}}
\def\me{m_e}
\def\te{T_e}
\def\ti{\tau_{\rm initial}}
\def\tci#1{n_e(#1) \sigma_T a(#1)}
\def\tr{\eta_r}
\def\dtr{\delta\eta_r}
\def\dd{\widetilde\Delta^{\rm Doppler}}
\def\dsw{\Delta^{\rm Sachs-Wolfe}}
\def\clsw{C_\ell^{\rm Sachs-Wolfe}}
\def\cldop{C_\ell^{\rm Doppler}}
\def\Dt{\widetilde{\Delta}}
\def\mut{\mu}
\def\vt{\widetilde v}
\def\hp{ {\bf \hat p}}
\def\sdv{S_{\delta v}}
\def\svv{S_{vv}}
\def\bvt{\widetilde{\bv}}
\def\delt{\widetilde{\delta_e}}
\def\cos{{\rm cos}}
\def\nn{\nonumber \\}
\def\bq{ {\bf q} }
\def\ba{ {\bf p} }
\def\bap{ {\bf p'} }
\def\bqp{ {\bf q'} }
\def\bp{ {\bf p} }
\def\bpp{ {\bf p'} }
\def\bk{ {\bf k} }
\def\bx{ {\bf x} }
\def\bv{ {\bf v} }
\def\qp{ p^{\mu}k_{\mu} }
\def\qpp { p^{\mu} k'_{\mu} }
\def\bgm{ {\bf \gamma} }
\def\bkp{ {\bf k'} }
\def\gq{ g(\bq)}
\def\gqp{ g(\bqp)}
\def\fp{ f(\bp)}
\def\h#1{ {\bf \hat #1}}
\def\fpp{ f(\bpp)}
\def\fz{f^{(\vec{0})}(p)}
\def\fpz{f^{(\vec{0})}(p')}
\def\f#1{f^{(#1)}(\bp)}
\def\fps#1{f^{(#1)}(\bpp)}
\def\dq{ {d^3\bq \over (2\pi)^32E(\bq)} }
\def\dqp{ {d^3\bqp \over (2\pi)^32E(\bqp)} }
\def\dpp{ {d^3\bpp \over (2\pi)^32E(\bpp)} }
\def\dtq{ {d^3\bq \over (2\pi)^3} }
\def\dtqp{ {d^3\bqp \over (2\pi)^3} }
\def\dtpp{ {d^3\bpp \over (2\pi)^3} }
\def\part#1;#2 {\partial#1 \over \partial#2}
\def\deriv#1;#2 {d#1 \over d#2}
\def\Done{\Delta^{(1)}}
\def\Dtwo{\widetilde\Delta^{(2)}}
\def\fone{f^{(1)}}
\def\ftwo{f^{(2)}}
\def\tg{T_\gamma}
\def\delpp{\delta(p-p')}
\def\delb{\delta_B}
\def\tc{\eta_0}
\def\DD{\langle|\Delta(k,\mu,\eta_0)|^2\rangle}
\def\DDL{\langle|\Delta(k=l/\tc,\mu)|^2\rangle}
\def\bkpp{{\bf k''}}
\def\kmkp{|\bk-\bkp|}
\def\kmkpsq{k^2+k'^2-2kk'x}
\def\tt{ \left({\tau' \over \tau_c}\right)}
\def\kt{ k\mu \tau_c}

%
%
%

\section{Introduction}

Recently, the study of supersymmetric field theories in Euclidean curved spaces has received considerable attention. 
In particular, the use of localization techniques has yielded a number of important results, including  
the exact computation  of the partition function,  expectation values  of Wilson loops and 't Hooft loops in  ${\cal{N}}=2$ 
theories on $\mathbb{S}^4$ \cite{P1,P2,P3,Gomis}. These calculations have been extended to the 
computation of the partition  function of supersymmetric gauge theories on other spaces such as  
$\mathbb{S}^3,~ \mathbb{S}^1\times\mathbb{S}^3 ,~\mathbb{S}^5$ and some deformed spheres 
(see e.g. \cite{hosomichi1,cheon,Jafferis2,hosomichi2,Kallen:2012va,hosomichi3,tera,Jafferis3,asano}). 
 Supersymmetric Yang-Mills (SYM) theories can be defined on these spaces.

A question of interest concerns the classification of all possible Euclidean curved spaces in various 
dimensions where one can have theories with rigid supersymmetry.
In four dimensions, one possible approach \cite{seiberg1} is to start with some supergravity theory coupled to 
matter multiplets in the off-shell formalism.
The idea is then to give backgrounds values to the gravity multiplet and to the auxiliary fields that preserve some 
supersymmetry and
then take the limit where the Planck mass goes to infinity. This limit should be taken 
in a way  that the gravitational dynamics decouples and one is left with
a theory with rigid supersymmetry on a frozen curved space. This approach has been further 
developed in many 
interesting works (see e.g. \cite{Jia:2011hw,Samtleben:2012gy,seiberg2,liu,dumitrescu,honda, lee1,lee2}, and
\cite{Kuzenko,kuzenko2} for earlier studies of rigid superspace geometry).
A different interesting approach is in terms of a holomorphic embedding of the space at 
the boundary of an asymptotically AdS space 
\cite{zaf1,zaf2}. 

The approach based on an off-shell formulation of supergravity is limited to the very few examples where an off-shell formulation is known,
for example, $N=2$ four dimensional supergravity or minimal five-dimensional supergravity. 
In this paper we shall show that  also the on-shell formalism of supergravity can be used to determine 
the spaces for  theories with global supersymmetry.
The basic idea is as follows.
One starts with any (Euclidean) supergravity action in $d$ dimensions, 
give background values to the gravity multiplet and possibly to other multiplets.
The resulting action will be supersymmetric if supersymmetry transformations do not change these values. This 
requires that supersymmetry transformations on all  fermions  vanish.  In particular, the vanishing of the gravitino 
shift gives an equation for the Killing spinor on a specific gravitational background. 
If a solution for the Killing spinor exists,
then there is some remaining supersymmetry (barring certain subtleties that appear in  dimensions $d>5$ --discussed in sections 5 and 6). 
This is in principle enough for the problem  of classification of supersymmetric curved spaces studied in this paper.

Having identified a given supersymmetric space, the next problem concerns the determination of the desired field theory Lagrangian with global supersymmetry.
Although finding specific field theory Lagrangians  goes beyond the scope of this work, in section 3.3 we   briefly comment on the prescription that one would have to follow 
within the on-shell approach.
The decoupling of gravity must be done in the usual way by taking the limit where the Planck mass goes to infinity. Obtaining non-trivial
curved spaces require that, at the same time, background values of matter fields are sent to infinity in an appropriate way.
 As long as the limit is regular, one is left with a field theory on a curved space which, by construction, 
 has global supersymmetry.

 In our quest for the classification of supersymmetric curved spaces, we will follow two different ways. The first one is based
 entirely on group theory. In fact, group theory already gives some model independent results in Poincar\'e supersymmetry.
 We may recall for example that group theory arguments restrict the maximal number of  spacetime dimensions
 to eleven 
 for a supersymmetric theory with one time direction and a single graviton \cite{nahm}. Similarly,  
 all manifest supersymmetries in different dimensions, including
those with conformal or de Sitter space-time symmetry, have been determined and, in particular, all possible 
simple supersymmetries have been classified \cite{nahm}. This gives us the possibility of 
identifying group theoretically all possible spaces admitting as isometry groups the 
bosonic part of the allowed supersymmetry groups. 
The second way of finding supersymmetric spaces is based on on-shell supergravity as explained above.

In the next section, we begin by studying a possible classification of supersymmetric spaces in various dimensions 
based on group theory arguments. In section 3 and  section 4 we discuss the 4D supergravity and the $N=2$ 5d 
supergravity. In section 5 the 6d $F(4)$ supergravity \cite{romans} is considered and in section 6 we generally 
comment 
 on supersymmetry on   $d>5$ spaces.  

\section{Group Theoretic Approach}

Supersymmetry generators  form a superalgebra \cite{kac,sorba,sorba2}. 
The latter has a graded $\mathbb{Z}_2$ structure which splits its generators 
into even and odd parts. The even generators form a classical algebra whereas the odd part transforms under some 
 representation of the even 
part. In supersymmetry algebra the odd part  is in the spinorial 
representation of its even (bosonic)  part. In fact, all possible simple supersymmetry algebras have been classified long 
ago by Nahm \cite{nahm}.  
By splitting the superalgebra ${\cal{G}}$ in the even $G_0$ and odd  $G_1$ parts as  
 ${\cal{G}}=G_0\oplus G_1$ 
with generators of $G_1$  
transforming in the $R$ representation of $G_0$, the possible superalgebras (in the Euclidean regime) are 
the following:
 

\begin{flalign}
&{\rm{\bf I.}}~~~ {\cal{G}}=F(4),~~~G=SO(6,1)\oplus SU(2),  ~~~R=(8,2)\ . &
 \end{flalign}
This case can describe a supersymmetric theory on the hyperbolic space $\mathbb{H}^6$ or a superconformal theory 
on  $\mathbb{S}^5$. 

\begin{flalign}
&{\rm{\bf II.}}~~ {\cal{G}}=su(4|N),~~~ G=SO(6)\oplus U(N), ~~~~R=(4,N)+(\overline{4},\overline{N})\ , ~~N\neq 4 \ ,& \nonumber\\
  &\phantom{{\rm{VIII.}}~~ }{\cal{G}}=su(4|4),~~~G=SO(6)\oplus SU(4), ~~~~R=(4,4)+(\bar{4},\bar{4}) \ , & 
\end{flalign}
for a supersymmetric theory on the round $\mathbb{S}^5$.

\begin{flalign}
 &{\rm{\bf III.}}~~~ {\cal{G}}=su^\ast(4|2N) \footnotemark \ ,~~~G=SO(5,1)\oplus U(2N)\ ,  
 ~~~R=(4,2N)+(\bar{4},\overline{2N})\ , &
\end{flalign}
\footnotetext{This case is missing from \cite{nahm} but appears in \cite{berezin} as $u_E(4,N)$.}
 for a superconformal theory on round $\mathbb{S}^4$ or 
a supersymmetric theory on the hyperbolic space $\mathbb{H}^5$.

 \begin{flalign}
&{\rm{\bf IV.}}~~~{\cal{G}}=osp(2|4),~~~G=SO(5)\oplus U(1), ~~~R=4+\bar{4}\ ,&
\end{flalign}
for a supersymmetric theory on the round $\mathbb{S}^4$. 
 
\begin{flalign}
&{\rm{\bf V.}}~~~ {\cal{G}}=osp(2|2,2), ~~~G=SO(4,1)\oplus U(1),  ~~~R=4+\bar{4}\ ,  &
\end{flalign}
for a superconformal theory on round $\mathbb{S}^3$  or 
a supersymmetric theory on the hyperbolic space $\mathbb{H}^4$. 

\begin{flalign}
&{\rm{\bf VI.}}~~~ {\cal{G}}=su(2|N)\oplus su(2|N), ~~~G=SO(4)\oplus U(N)^2,  ~~~R=(2,1,N,1)+(1,2,1,N), \quad N\neq 2 \ ,  &
\end{flalign}
\begin{flalign}
&{\rm or}\ \ \ \ {\cal{G}}=su(2|2)\oplus su(2|2), ~~~G=SO(4)\oplus SU(2)^2,  ~~~R=(2,2,1,1)+(1,1,2,2)\ , &
\end{flalign}
for a supersymmetric theory on round $\mathbb{S}^3$. 

 \begin{flalign}
&{\rm{\bf VII.}}~~~{\cal{G}}=osp(3,1),~~~G=SO(3,1), ~~~R=(2,1)+(1,2) \ ,&
\end{flalign}
for a superconformal theory on round $\mathbb{S}^2$ or 
a supersymmetric theory on the hyperbolic space $\mathbb{H}^3$. 

\begin{flalign}
 &{\rm{\bf VIII.}}~~~ {\cal{G}}=su(2|N), ~~~G=SO(3)\oplus U(N),  ~~~R=(2,N)+(2,\bar{N}) \ ,&\nonumber \\
&\phantom{{\rm{\bf VII.}}}~~~ {\cal{G}}=su(2|2), ~~~G=SO(3)\oplus SU(2),  ~~~R=(2,2)+(2,2) \ ,&
 \end{flalign}
for a supersymmetric theory on the round $\mathbb{S}^2$.

\begin{flalign}
&{\rm{\bf IX.}}~~~ {\cal{G}}=su(1,1|N), ~~~G=SO(2,1)\oplus U(N),  ~~~R=(2,N)+(2,\bar{N}) \ ,&\nonumber \\
&\phantom{{\rm{\bf IX.}}}~~~ {\cal{G}}=su(1,1|2), ~~~G=SO(2,1)\oplus SU(2),  ~~~R=(2,2)+(2,2) \ ,&
 \end{flalign}
\begin{flalign}
&{\rm{\bf X.}}~~~ {\cal{G}}=osp(N|2), ~~~G=SO(2,1)\oplus SO(N),  ~~~R=(2,N) \ ,&
 \end{flalign}
 and 
\begin{flalign}
&{\rm{\bf XI.}}~~~ {\cal{G}}=osp(4|2,a), ~~~G=SO(2,1)\oplus O(4),  ~~~R=(2,4)_a \ ,&
 \end{flalign}
for  supersymmetric theories on the hyperboloid $\mathbb{H}^2$, or  superconformal theories on $\mathbb{S}^1$.

\begin{flalign}
&{\rm{\bf XII.}}~~~ {\cal{G}}=F(4), ~~~G=SO(2,1)\oplus SO(7),  ~~~R=(2,8)\ ,&
 \end{flalign}

\begin{flalign}
&{\rm{\bf XIII.}}~~~ {\cal{G}}=G(3), ~~~G=SO(2,1)\oplus G_2,  ~~~R=(2,7)\ ,&
\end{flalign}
for  supersymmetric theories on the hyperboloid $\mathbb{H}^2$, or  superconformal theories on $\mathbb{S}^1$. 
 
\begin{flalign}
&{\rm{\bf XIV.}}~~~ {\cal{G}}=osp(2|2N), ~~~G=SO(2)\oplus Sp(2N),  ~~~R=2N\oplus 2N\ ,&
\end{flalign}
for a supersymmetric theory on $\mathbb{S}^1$.

\medskip

 We have collected the above in Table \ref{table1} and represented the cases of simply connected, 
 maximally symmetric spaces admitting supersymmetric theories.
It should be noted that round spheres $\mathbb{S}^d$ with $d>5$, and hyperboloids $\mathbb{H}^d$ with $d>6$,
are not allowed by the Nahm classification.
We will further comment on this in section 5 and section 6.

Product spaces are also possible. They correspond to non-simple superalgebras and some 
interesting cases are presented in 
Table \ref{table2}.  Note that, as we will see in the next sections,   only products with 
 $\mathbb{S}^1$ factors  preserve conformal flatness.
 Nevertheless, some more general direct product spaces are also compatible with the group theory classification. For example:
 \begin{itemize}
 
 \item $S^2\times S^1\times S^1$.  This has $SO(3)\times U(1)\times U(1)$ isometries. It can be
 embedded in a superalgebra ${\cal{G}}=su(2|1)\oplus osp(2|2)$.

\item $S^2\times S^1\times S^1$. This has $SO(4)\times U(1)\times U(1)$, which exactly matches the bosonic symmetries of case {\bf VI} with $N=1$, 
${\cal{G}}=su(2|1)\oplus su(2|1)$.

\end{itemize}

\noindent Both cases also satisfy the condition that the odd part transforms in the spinorial representation
of the even part.

\bigskip

\begin{table}[!h]
\centering
\begin{tabular}{|c|c|c|c|c|}
\hline\hline
 ${\cal{G}}$&$G_0$&R&SUSY&SC\\
 \hline\hline
$osp(6,1|2)$&$SO(6,1)\oplus SU(2)$&(8,2)&$\mathbb{H}^6$&$\mathbb{S}^5$\\
\hline
$su(4|N)$&$SO(6)\oplus U(N)$ &$(4,N)+(\bar{4},\bar{N}), ~~N\neq 4$&$\mathbb{S}^5$&\\
 \hline
$su(4|4)$&$SO(6)\oplus SU(4)$ &$(4,4)+(\bar{4},\bar{4})$&$\mathbb{S}^5$&\\
\hline
$su^\ast(4|2N)$&$SO(5,1)\oplus U(2N)$&$(4,2N)+(\bar{4},\bar{2N})$& $\mathbb{H}^5$ &$\mathbb{S}^4$\\
\hline
$osp(2|4)$&$SO(5)\oplus U(1)$&$4+\bar{4}$&$\mathbb{S}^4$&\\
\hline 
$osp(2|2,2)$&$SO(4,1)\oplus U(1)$ & $4+\bar{4}$ & $\mathbb{H}^4 $ &$\mathbb{S}^3$\\
\hline
$su(2|N)\oplus su(2|N)$& $SO(4)\oplus U(N)^2$ & $(2,\bar N,1,1)+(1,1,2,\bar N)$ & $\mathbb{S}^3 $ &\\
\hline
$su(2|2)\oplus su(2|2)$& $SO(4)\oplus SU(2)^2$ & $(2,2,1,1)+(1,1,2,2)$ & $\mathbb{S}^3 $ &\\
\hline
$su(2|N)$&$SO(3)\oplus U(N)$&$(2,N)+(2,\bar{N})$& $\mathbb{S}^2$&\\
 \hline
$osp(3|2)$&$SO(3)\oplus SU(2)$&$(2,2)+(2,\bar{2})$& $\mathbb{S}^2$&\\
\hline
$osp(3,1)$&$SO(3,1)$&$(2,1)+(1,2)$& $\mathbb{H}^3$ & $\mathbb{S}^2$\\
\hline
$su(1,1|N)$&$SO(2,1)\oplus U(N)$&$(2,N)+(2,\bar{N})$ &$\mathbb{H}^2$&$\mathbb{S}^1$\\
 \hline
$su(1,1|2)$&$SO(2,1)\oplus SU(2)$&$(2,2)+(2,\bar{2})$ &$\mathbb{H}^2$&$\mathbb{S}^1$\\
\hline
$osp(N|2)$&$SO(2,1)\oplus SO(N)$&$(2,N)$&$\mathbb{H}^2$&$\mathbb{S}^1$\\
 \hline
$osp(4|2,a)$&$SO(2,1)\oplus O(4)$&$(2,4)_a$& $\mathbb{H}^2$&$\mathbb{S}^1$\\
\hline
$F(4)$&$SO(2,1)\oplus SO(7)$&  $(2,8)$&$\mathbb{H}^2$&$\mathbb{S}^1$\\
 \hline
$G(3)$&$SO(2,1)\oplus G_2$&  $(2,7)$&$\mathbb{H}^2$&$\mathbb{S}^1$\\
\hline
$osp(2|2N)$& $SO(2)\oplus Sp(2N)$&$2N\oplus 2N$&$\mathbb{S}^1$&\\
 \hline\hline
\end{tabular}
\caption{The superalgebras for maximally symmetric spaces.}
\label{table1}

\end{table}

\begin{table}[!h]
\centering
\begin{tabular}{|c|c|c|}
\hline\hline
Dimension&Background&Superalgebra\\
\hline\hline
\multirow{2}{*}{2}&\multirow{2}{*}{$\mathbb{S}^1\times \mathbb{S}^1$}&${\cal{G}}_1\oplus {\cal{G}}_1$\\
\cline{3-3}
& & $osp(2|2N)\oplus {\cal{G}}_2$ \\
\hline
\multirow{2}{*}{3}&\multirow{2}{*}{$\mathbb{S}^1\times \mathbb{S}^2$}&$osp(3,1)\oplus {\cal{G}}_1$\\
\cline{3-3}
& & $su(2|N)\oplus {\cal{G}}_2$ \\
\hline
\multirow{2}{*}{4}&\multirow{2}{*}{$\mathbb{S}^1\times \mathbb{S}^3$}&$osp(2|2,2)\oplus {\cal{G}}_1$\\
\cline{3-3}
& & $su(2|N)\oplus su(2|N)\oplus {\cal{G}}_2$\\
\hline
\multirow{2}{*}{5}&\multirow{2}{*}{$\mathbb{S}^1\times \mathbb{S}^4$}&$su^\ast(4|2N)\oplus {\cal{G}}_1$\\
\cline{3-3}
&&$osp(2|4)\oplus {\cal{G}}_2$\\
\hline
\multirow{2}{*}{6}&\multirow{2}{*}{$\mathbb{S}^1\times \mathbb{S}^5$}&$osp(6,1|2)\oplus {\cal{G}}_1$ \\
\cline{3-3}
&&$su(4|N)\oplus {\cal{G}}_2$ \\
\hline
\end{tabular}

\caption{Some interesting product spaces and their corresponding superalgebra are given by the (non-exhaustive) 
list above.  ${\cal{G}}_1$ is one of 
$\{su(2|N),osp(N|2),osp(4|2,a),F(4),G(3)\}$ and 
${\cal{G}}_2$ is one of $(su(2|N),osp(N|2),osp(4|2,a),F(4),G(3), osp(2|2N))$.}
\label{table2}
\end{table}

\medskip
\bigskip

\subsection{Ellipsoids}

In a similar manner one can also have superalgebras on ellipsoids in several dimensions. They are generally described
by the equation
\be
\sum_{\mu=1}^d \frac{x_\mu^2}{\ell_\mu^2}=1\ .
\ee
The superalgebra classification 
is  determined in terms of the bosonic isometries. Examples are given below:

\begin{itemize}

\item Ellipsoids preserving only $U(1)$ isometries allow for  superalgebras of the form
$$
{\cal{G}}=osp(2|2N)\oplus ...\oplus osp(2|2N)\ ,
$$
i.e. direct sums of type XIV superalgebras.

\item Ellipsoids preserving an $SO(3)$ isometry and a $U(1)$ isometry  allow for  superalgebras of the form
$$
{\cal{G}}=osp(3|N)\oplus osp(2|2N) \ ,
$$
i.e. it is a direct sum of a type VIII and type XIV superalgebras.

\item Ellipsoids preserving an $SO(4)$ isometry and a $U(1)$ isometry allow for  superalgebras of the form
$$
{\cal{G}}=sl(2|N)\oplus sl(2|N)\oplus  osp(2|2N)\ . 
$$

\item Ellipsoids preserving an $SO(5)$ isometry and a $U(1)$ isometry allow for  superalgebras of the form
$$
{\cal{G}}=osp(2|4)\oplus osp(2|2N) \ .
$$

\end{itemize}


\subsection{Orbifolds}

There is  another class of manifold which admits supersymmetry and is connected to the  cases 
I--XIV described  above. Their construction is as follows. Consider a manifold $N$ with isometry group $G$ 
and let $\Gamma\subset G$ a 
discrete subgroup of the latter freely acting on $N$. Then the space $M=N/\Gamma$ is non-singular and corresponds to global
identifications of $N$. Therefore, the isometries of $M$ will be different from those of $N$ leading to 
different supersymmetry algebra supported by $M$. To be precise, let us consider in particular
$\mathbb{S}^3$, as the other cases are quite similar. 

The group of orientation-preserving isometries of $\mathbb{S}^3$ is $SO(4)$. The quotient of
$SO(4)$ by its center $\{\pm I\}$ is isomorphic to $SO(3) \times SO(3)$, therefore
a finite subgroup $G$ of $SO(4)$ gives rise to two finite subgroups $G_L$ and $G_R$  of $SO(3)$. 
These considerations specify the possible finite subgroups of $\mathbb{S}^3$ to be
\be
\Gamma=\mathbb{Z}_n\, ,~~ D^\ast_{4n}\, ,~~ T^\ast_{24}\, , ~~ O^\ast_{48}, ,~~ I^\ast_{120}\, , 
\ee
i.e.,  the finite cyclic groups $\mathbb{Z}_n$, the binary dihedral groups $D_{4n}$ of order 4n, 
and the binary tetrahedral, octahedral, 
and icosahedral groups $T^\ast_{24}$, $O^\ast_{48}$, and $I^\ast_{120}$, of orders 24, 48, and 120, 
respectively. If $G$ acts freely on 
$\mathbb{S}^3$, then, say, $G_L$ must be 
cyclic, and $G_R$ can then be described as being of cyclic, dihedral, tetrahedral, octahedral, or icosahedral type,
according to the type of $G_R$. The groups of cyclic type are cyclic, and the corresponding 3-manifolds are the lens 
spaces $L(m,n)$ defined by the identification 
\be
(z_1,z_2)\equiv (e^{2\pi i/m}z_1,e^{2\pi i n/m} z_2)
\ee
of the coordinates of $(z_1,z_2)$ of $\mathbb{C}^2$ where  $\mathbb{S}^3$ is embedded.  
 The isometry groups of $\mathbb{S}^3/\Gamma$ have been calculated in  \cite{mac}. Here we just mention that for 
lens spaces we have for example that 
\be
\mathtt{isom}\big{(}\mathbb{S}^3/T^\ast_{24}\big{)}=SO(3)\times \mathbb{Z}_2\, , ~~~~
\mathtt{isom}\big{(}\mathbb{S}^3/T^\ast_{24}\times \mathbb{Z}_2\big{)}=SO(2)\times \mathbb{Z}_2
\ee
where $\mathtt{isom}(M)$ is the isometry group of $M$. Similarly, for the lens spaces 
$L(m,1)=\mathbb{S}^3/Z_m$ and  $L(m,n)$  we have  
\bea
&&\mathtt{isom}\big{(}\mathbb{S}^3/Z_m\big{)}= SO(3)\times U(1)\, ,\qquad m>2 \nonumber \\
&&\mathtt{isom}\big{(}L(m,n)\big{)}= U(1)\times U(1)\qquad ~~ m,\,  (n^2-1)/m\, , ~\mbox{even} 
\eea
These isometry groups will then be  the even part of the supersymmetry algebra on $\mathbb{S}^3/\Gamma$. For example, 
let us take the case of $\mathbb{S}^3/Z_m$. Starting with  the round $\mathbb{S}^3$, we may consider the simplest case 
of an $osp(4|2n)$ superalgebra with bosonic subgroup $G_{0}=SU(2)\times SU(2)\times Sp(2n)$. By embedding the $\mathbb{Z}_m$
in the $U(1)$ subgroup of the second $SU(2)$, the isometry group is broken down to $SU(2)\times U(1)$. Similarly,
the fermionic generators are in the $(2,1,2n)+(1,2,2n)$ representation of $G_0$ and 
only the $(2,1,2n)$ part survives  the modding by 
$\mathbb{Z}_m$.  Therefore, the supersymmetry algebra for the $\mathbb{S}^3$ has been reduced to 
$osp(3|2n)$. Similar considerations apply to the other cases and in higher dimensional spheres as well. 
The  partition function for super Yang-Mills theory on $\mathbb{R}\times \mathbb{S}^3/\mathbb{Z}_m$ has been recently computed
in \cite{asano}.

\section{Supersymmetric 4d spaces from matter superfields coupled to supergravity}

\subsection{$N=1$ supergravity}

In \cite{seiberg1,seiberg2,liu}, supersymmetric spaces are obtained by giving background values to auxiliary fields. In this approach,
the auxiliary fields are not required to satisfy the equations of motion.
The curved geometries are then supported by the background auxiliary fields.
It is interesting to see how the different possible curved spaces are realized in the on-shell formalism,
where auxiliary fields have already been eliminated by their equations of motion.
In this approach one has to give background values to some dynamic fields.

In this subsection we first discuss in detail the case of  old minimal  $N=1$ supergravity with Euclidean signature
coupled to chiral superfields, and then consider  the addition of vector multiplets.
{} Supersymmetry requires that the supersymmetry transformations on  fermions vanish on a given bosonic background.
In the Euclidean theory the  left and right handed components of  a fermion  $P_L\psi$ and $ P_R\psi$ will be independent. We will denote them by $\psi $ and $\bar\psi $.
The supersymmetry transformations for the  left and right handed gravitino are (see (18.22) in \cite{freedman})
\bea
&& 
\delta \psi_\mu = \nabla_\mu \xi +  i A_\mu \xi +\frac{1}{2}  B \Gamma_\mu \bar\xi\ ,
\\
&& 
 \delta \bar \psi_\mu =\nabla_\mu \bar\xi -  iA_\mu \bar \xi+ \frac{1}{2} \bar B \Gamma_\mu \xi\ ,
\eea
where
\be
A_\mu\equiv -{i \kappa^2\over 4} \big(  K_\alpha \p_\mu \phi^\alpha -  K_{\bar\alpha} \p_\mu \bar \phi^{\bar \alpha}\big)\ ,\qquad B\equiv {\kappa^2}\ e^{\kappa^2 K/2}\ W\ ,
\label{daz}
\ee
and
\be
\nabla _\mu\xi =\big(\partial_\mu +{1\over 4}\omega_\mu^{ab} \Gamma_{ab}\big)\xi
\ ,\qquad 
\Gamma_{ab}={1\over 2}\big(\Gamma_a \Gamma_b-\Gamma_b\Gamma_a\big)\ .
\nonumber
\ee
As usual, $W=W(\phi^\alpha)$, $\alpha=1,...,n_c$, denotes the superpotential and $K=K(\phi^\alpha,\bar \phi^\alpha)$ is the K\"ahler potential. $\mu, \nu$ and $a,b$  respectively denote curved space and tangent space indices.
We shall follow the conventions of \cite{freedman} for spinors and Dirac $\Gamma $ matrices.
We also need to set to zero the supersymmetry variations of the fermions of the chiral multiplets, $\delta\chi^\alpha =\delta\bar\chi^\alpha=0$. They will be discussed below.

Consider the equations $\delta \psi_\mu =\delta \bar \psi_\mu =0$. The integrability condition for the equation $\delta \psi_\mu =0$ gives
\bea
0 &=& \big[ \nabla_\nu,\nabla_\mu \big] \xi +iF_{\nu \mu}\ \xi +iA_\mu \nabla_\nu\xi -iA_\nu \nabla_\mu\xi 
\nonumber\\
&+&{1\over 2} \left( \nabla_\nu B \Gamma_\mu\bar\xi - \nabla_\mu B \Gamma_\nu\bar\xi +B\Gamma_\mu\nabla_\nu \bar\xi -B\Gamma_\nu\nabla_\mu \bar\xi \right)\ .
\eea
Using the equations for $\xi,\ \bar\xi$, this becomes
\bea
 0 &=& {1\over 4} R_{\nu\mu ab}\Gamma^{ab} \xi +
iF_{\nu\mu}\ \xi  -{1\over 2} B\bar B \Gamma_{\mu\nu} \xi 
\nonumber\\
&-&{1\over 2}  \left[ \Gamma_\nu\left( \nabla_\mu B +2i B A_\mu \right)\bar{\xi} - 
\Gamma_\mu\left( \nabla_\nu B +2i B A_\nu \right)\bar{\xi}\right]\ .
\eea
Assuming maximal supersymmetry, we get the conditions
\be
F_{\mu\nu}=0\ ,\qquad \nabla_\mu B =-2i B A_\mu \ ,\qquad \nabla_\mu \bar B = 2i \bar B A_\mu \ ,
\label{condin}
\ee
\be
R_{\nu\mu ab}\Gamma^{ab} \xi = 2 B\bar B \Gamma_{\mu\nu} \xi\ .
\label{ara}
\ee
This last equation implies that
\be
R_{\mu\nu \rho\sigma}= - B\bar B \ \big( g_{\mu\rho}g_{\nu\sigma}- g_{\mu\sigma}g_{\nu\rho}\big) \ ,
\ee
i.e. the space is locally isometric to a maximally symmetric space.
Therefore the Weyl tensor  $W_{\mu \nu\rho\sigma }$ vanishes and the space is conformally flat.
We have
\be
W_{\mu \nu \rho\sigma}=0\ ,\qquad R_{\mu\nu}= - 3 B\bar B \ g_{mn}\ .
\label{afa}
\ee
Let us now examine the conditions (\ref{condin}) in detail. 
One solution is
\be
B=0\ ,\qquad A_\mu={\rm arbitrary}\ .
\ee
This gives flat space-time.

If $B\neq 0$, then (\ref{condin}) is solved by an $A_\mu$ of the form, $A_\mu=\nabla_\mu \Lambda $. However, from the definition  (\ref{daz}) of $A_\mu$,
the integrability condition then leads to $K_{\alpha\bar \beta}d\phi^\alpha\wedge d\phi^{\bar \beta} =0$, which implies $\phi^\alpha =\phi^\alpha_0={\rm constant}$, hence $A_\mu=0$.
Therefore the unique solution is
\be
A_\mu=0\ ,\qquad B=B_0={\rm const}\ .
\ee

{}  We still need to check the equation for the fermions of the chiral multiplets.
For constant scalars, this gives
\bea
\delta\chi^\alpha  &=& -{1\over\sqrt{2}} e^{\kappa^2K\over2} (K^{\alpha\bar \beta}\bar\nabla_{\bar \beta} \bar W)\ \xi=0\ ,
\nonumber\\
\bar \delta\chi^{\bar \alpha} &=& -{1\over\sqrt{2}} e^{\kappa^2K\over2} (K^{\bar \alpha \beta}\nabla_{ \beta}  W)\ \bar \xi=0\ .
\eea
 Thus we get the following condition for the constant values $\phi_0,\ \bar \phi_0$ of scalar fields:
\be
\nabla_{ \alpha}  W\bigg|_{\phi_0}=\bar\nabla_{\bar \alpha} \bar W\bigg|_{\phi_0} = 0\ .
\ee

\medskip

Let us now consider the addition of vector multiplets. 
We now need to impose, in addition, that the supersymmetry transformation of the gauginos vanishes.
Assuming constant values for the vector bosons, this gives the extra condition
\be
\delta \lambda^A = {i\over 2} \gamma_5 ({\rm Re} f)^{-1\ AB}{\cal P}_B \xi =0\ ,
\ee
where $f_{AB}(\phi)$ are the holomorphic functions determining the gauge multiplet kinetic terms
and ${\cal P}_A(\phi,\bar\phi)$ are the Killing potentials.
This implies
\be
{\cal P}_A(\phi_0,\bar \phi_0)=0\ .
\ee
This is an extra condition on the constant background values $\phi^\alpha_0,\ \bar\phi^\alpha_0$.
 The gravitino transformation is not changed.
Therefore, turning background values for  vector bosons (Wilson lines) does not generate new supersymmetric spaces.

\medskip

Thus we find Einstein-Weyl spaces of negative curvature.  This  implies that the space is locally isometric to $\mathbb{H}^4$. 
Supersymmetric theories on a positive curved space $\mathbb{S}^4$ can be obtained by relaxing the condition
that $\bar B$ is the complex conjugate of $B$. As discussed in an analogous context in \cite{seiberg1},
this leads to a Lagrangian which is not real. 
The resulting Euclidean theory is not reflection positive and does not correspond to any unitary theory with Lorentzian signature.
This is not surprising, since general supersymmetric theories cannot be put on $dS_4$. 
An exception occurs when the theory is superconformal; 
then $dS_4$ is admitted, as this space is conformal to Minkowski space. In this case the Lagrangian becomes real.

\medskip

It should be noted that, for superconformal theories, more general spaces are allowed.
Indeed, they can be formulated in any space which is conformal to $\mathbb{H}^4$ (or, equivalently, to $\mathbb{S}^4$).
This gives more options, in particular, spaces of the form $X\times \mathbb{S}^1$, where $X$ is locally isometric to a maximally symmetric space.
We next show that spaces of the form $X\times \mathbb{S}^1$ can also be admitted in non-superconformal theories if a suitable Wilson line background field corresponding to an R-symmetry 
is turned on in the $\mathbb{S}^1$ direction.

\subsection{$N=2$ four-dimensional gauged supergravity}

Here we shall  show that theories with rigid supersymmetry can be formulated on spaces of the form $X\times \mathbb{S}^1$, where $X$ is locally isometric to a maximally symmetric space, 
by turning on suitable R-symmetry vector field components  in the $\mathbb{S}^1$ direction.

The mechanism can be implemented in any dimension $d\leq 6$, starting with a suitable gauged supergravity. Different dimensions need to be examined case by case, because
there are  important technical differences, in particular due to the 
different 
spinor representations.
In this section we begin by considering  the four-dimensional case. 

Our starting point is ${\cal N}=2$ gauged supergravity coupled to any number of vector and hyper multiplets.
For a detailed description of the theory we refer to \cite{freedman}.
The scalar manifold is a direct product of the special K\"ahler manifold of the scalars in the vector multiplets and the quaternionic K\"ahler manifold of the scalars in
the hypermultiplets.
The graviton multiplet contains the vierbein $e^a_\mu$, two gravitinos $\psi^i_\mu$ and the graviphoton $A_\mu^0$.
By turning on constant values for the complex scalar fields $z^\alpha ,\ \bar z^{\bar\alpha}$ of the
vector multiplet ($\alpha=1,...,n_V$),  constant values for the real scalars $q^u$ of the hypermultiplets ($u=1,...,4n_H$),
and constant values for  vector field components $A_\mu^I $ ($I=0,1,...,n_V$), the supersymmetry transformation law for the left and right handed
gravitinos  take the form (see \cite{freedman}, eq. (21.42)),
\be
\delta \psi_\mu^i = \nabla_\mu\xi^i - {i\over 2} {\cal A}_\mu \xi^i+ {\cal V}_{\mu \ j}^{\ i}\xi^j +{1\over 2} \kappa^2 \Gamma_\mu S^{ij}  \xi_j = 0\ ,
\ee
where
\be
{\cal A}_\mu=-\kappa^2 A_\mu^I P_I^0\ ,\qquad {\cal V}_{\mu}^{ij}=-{\kappa^2\over 2} A_\mu^I P_I^{ij}\ ,\qquad S^{ij}= P_{I}^{ij}\bar X^I\ ,
\ee
and $ P_I^{ij}(q)$ denote, as usual, moment maps on the quaternionic K\"ahler metric $g_{XY}$ 
of the hypermultiplet scalar manifold, and $P_I^0(z,\bar z)$ is the real moment map of the special K\"ahler manifold.
$S^{ij}$ is a  symmetric matrix which
depends on the constant background values for the scalars $q,\ \bar z$.
We recall that spinors are $SU(2)$ doublets and $SU(2)$ indices are lowered and raised by $\epsilon_{ij}$.

{} To illustrate the method, we again begin by looking for spaces that preserve
a maximum amount of supersymmetry.
To simplify the discussion in what follows we assume that ${\cal V}_{\mu \ j}^{\ i}=0$, since it is not needed to generate the relevant solutions.

The supersymmetry parameter $\xi_i$ is a symplectic Majorana spinor satisfying
\be
\xi_i=\xi^j\epsilon_{ji}\ ,\qquad \xi_i=(\xi^i)^C\ ,
\ee
where $\lambda^C$ denotes charge conjugation. Like in the $N=1$ case, in Euclidean space we must relax this condition and treat
$\xi^i $ and $\bar \xi^i\equiv (\xi^i)^C$ as independent.\footnote{Supersymmetry in Euclidean space is an old subject on its own \cite{euclidean}. 
In Euclidean space the charge conjugation matrix has imaginary eigenvalues and the
Majorana condition cannot be imposed, although there are alternative treatments (see e.g. \cite{VanNiew}).}

Thus we have two independent equations
\bea
&&\delta \psi_\mu^i = \nabla_\mu\xi^i - {i\over 2} {\cal A}_\mu \xi^i +{1\over 2} \kappa^2 \Gamma_\mu S^{ij} \bar \xi^j = 0\ ,
\nonumber\\
&&\delta \bar\psi_\mu^i = \nabla_\mu \bar \xi^i + {i\over 2} \bar {\cal A}_\mu \bar\xi^i +{1\over 2} \kappa^2 \Gamma_\mu \bar S^{ij} \xi^j = 0\ .
\label{cor}
\eea
Likewise, $S^{ij}$ and $\bar S^{ij}$ and  ${\cal A}_\mu$ and $\bar {\cal A}_\mu$ will be treated as independent.
This doubling of some boson degrees of freedom in Euclidean space seems to be natural
in view of the doubling of some fermion degrees of freedom, although it is not clear
how this should be done consistently in the full theory (see e.g. \cite{liu} for a recent discussion).

The supersymmetry variations of other fermions will be discussed below.

As in the $N=1$ case, we first look for spaces with maximal supersymmetry.
The integrability condition of (\ref{cor}) implies that
\be
F_{\mu\nu}=0\ .
\ee
which is automatically satisfied for constant $ {\cal A}_\mu$.
We are left with
\bea
0 &=& [\nabla_\nu,\nabla_\mu]\xi^i -{\kappa^4\over 2} S^{ij}\bar S^{jk} \Gamma_{\mu\nu} \xi^k +{i\kappa^2\over 2}({\cal A}_\mu\Gamma_\nu-{\cal A}_\nu\Gamma_\mu) S^{ij}\bar\xi^j\ ,
\nonumber\\
0 &=& [\nabla_\nu,\nabla_\mu]\bar \xi^i -{\kappa^4\over 2} \bar S^{ij} S^{jk} \Gamma_{\mu\nu}\bar \xi^k  -{i\kappa^2\over 2}(\bar {\cal A}_\mu\Gamma_\nu-\bar {\cal A}_\nu\Gamma_\mu)\bar S^{ij}\xi^j\ .
\eea
We will assume that $[S, \bar S]=0$.
Let us now consider the supersymmetry variation of the gauginos and hyperinos.
Analogously to the   $N=1$ case, these lead to algebraic constraints
on the constant values of the scalar fields. In the notation of \cite{freedman}
\be
\bar W_{\bar\beta ji}(z,\bar z,q)=  W_{\beta ji}(z,\bar z, q)=0\ ,\qquad \bar N^A_i(z,\bar z, q)= N^A_i(z,\bar z,q)=0\ .
\label{maaa}
\ee
Solving these equations explicitly requires specifying the model. Note that these points correspond to fixed points of the scalar manifold in supersymmetric flows.
The standard relation that connects the scalar potential $V$ to fermion shifts now gives the identity
\be
-3\kappa^2 S^{ik}\bar S_{jk}=\delta^i_{\, j}\ V
\ee
If ${\cal A}_\mu=0$, the gravitino equation can be solved without any restriction on the spinors.
It implies 
\be
W_{\mu\nu \rho\sigma}= 0\ ,\qquad R_{\mu\nu}= \kappa^2 V  g_{\mu\nu}\ .
\ee
i.e., the space is Einstein with vanishing Weyl tensor. 
The choices $V<0$ and $V>0 $ respectively give spaces locally isometric to $\mathbb{H}^4$ and $\mathbb{S}^4$.

\medskip

Consider now a reducible space of the form $ X_3\times \mathbb{S}^1$. As this space has non-trivial homotopy group $\pi_1(\mathbb{S}^1)=\mathbb{Z}$ one can turn on
Wilson lines. Then one can find the following solution.
We turn on  constant components ${\cal A}_4$, $\bar {\cal A}_4$ and ${\cal A}_{\hat \mu}=\bar {\cal A}_{\hat\mu}=0$, $\hat\mu=1,2,3$, and constant scalars.
The vanishing of the gaugino and hyperino variations again imply the conditions (\ref{maaa}).
Consider now the gravitino variation.
It is  convenient to  consider a spinor basis $\xi $, $\bar \xi $ where $\Gamma_4$ is diagonal.
We demand, 
\be
\p_4\xi_ i= \p_4\bar \xi_i = 0\ .
\ee
The fourth component of the conformal Killing spinor equation (\ref{cor}) then reads
\be
i {\cal A}_4 \xi^i = \kappa^2 \Gamma_4 S^{ij}\bar \xi^j\ , \qquad i \bar {\cal A}_4 \bar \xi^i = -\kappa^2 \Gamma_4 \bar S^{ij} \xi^j\ .
\label{ret}
\ee
Equation  (\ref{ret}) reduces the number of supersymmetries by a factor of 1/2. In particular, if 
the first equation is solved for the spinors  with $\Gamma_4=1$, the spinors with $\Gamma_4=-1$
must be set to zero.
%
Combining both equations, we find
\be
{\cal A}_4 \bar {\cal A}_4  = -\frac{\kappa^2}{3} \, V  \ ,
\ee
with no further restriction on the spinors.
Substituting into the remaining equations, we find
\be
 \nabla_{\hat\mu} \xi^i = \frac{i}{2}  {\cal A}_4 \Gamma_4 \Gamma_{\hat\mu}\xi ^i \ ,\qquad  \nabla_{\hat \mu} \bar \xi^i =- \frac{i}{2}  \bar {\cal A}_4 \Gamma_4 \Gamma_{\hat\mu}\bar \xi ^i \ .
\ee
The integrability conditions of these equations imply that
\be
W_{\hat\mu \hat\nu \hat\rho \hat\sigma}(X)=0\ ,\qquad R_{\hat\mu\hat\nu} = \frac{2\kappa^2}{3}\ V \ g_{\hat\mu\hat\nu}\ ,
\ee
\be
{\cal A}_4^2 =\bar {\cal A}_4^2 =-\frac{\kappa^2}{3} \, V  \ .
\ee
The Einstein-Weyl condition implies that $ X_3$ is locally isometric to  a maximally symmetric space.
Therefore we get spaces $X_3\times \mathbb{S}^1$ by turning on a Wilson line.
According to  the sign of $V $, we find spaces locally isometric to $ \mathbb{H}^3\times \mathbb{S}^1$ or $\mathbb{S}^3\times \mathbb{S}^1$.

\medskip

Finally, note that the backgrounds discussed so far do not allow for supersymmetric theories on ellipsoids.
 Killing spinors on ellipsoids can be obtained by turning on suitable $SU(2)_R$ gauge fields and tensor fields \cite{hosomichi3}.
This suggests that it might be possible to obtain theories with rigid supersymmetry on ellipsoids from on-shell $N=4$ gauged supergravity.
It would  be  interesting to construct the explicit solution  of the Killing spinor equations in this way.

\subsection{Comments on the construction of supersymmetric Lagrangians}

The off-shell treatment in supergravity is limited to few examples 
such as $N=2$ four dimensional supergravity or minimal five-dimensional supergravity. 
Therefore, it is important to understand
the construction of supersymmetric Lagrangians within the on-shell formalism.

Consider first the $N=1$ four-dimensional case discussed in section 3.1. 
One starts with supergravity coupled to the desired number of chiral and vector multiplets, having the desired interactions. 
One then adds   an extra set of self-interacting chiral multiplets with a convenient superpotential $W$, whose r\^ ole will be to provide the background values $\phi_0,\ \bar\phi_0$,
to support, for example, $\mathbb{S}^4$ spaces. They couple to gravity, but they do not couple to the chiral and vector multiplets of the physical theory that one wishes to study.
By construction, the Lagrangian is supersymmetric, since the background values $\phi_0,\ \bar\phi_0$ solve the conditions for
supersymmetry.
Then one takes the limit where the Planck mass $M_P $ goes to infinity, i.e. $\kappa\to 0$ by first rescaling $\psi_\mu\to \kappa\psi_\mu $ and
with fixed $\kappa \phi_0 $, $\kappa \bar \phi_0$.

More generally, to construct a supersymmetric theory on a curved $d$-dimensional space, 
one shall start with a  suitable $d$-dimensional gauged supergravity with $N\leq 4$ supersymmetries, couple it
to the desired matter multiplets, plus additional free matter multiplets whose r\^ ole is to provide  background
that support the curved supersymmetric space.
The detailed construction of Lagrangians in specific models is beyond the scope of this paper, which is motivated by the problem of classification.

\section{Supersymmetric spaces in five dimensions}


We shall consider $N=2$   gauged supergravity coupled to $n_V$  vector multiplets and $n_H$ hypermultiplets.
One may also consider adding tensor multiplets, though for our purposes this is unnecessary.
We recall that the fields of the $N = 2$ supergravity multiplet are the f\"unfbein $e_\mu^a$, two gravitini $\psi_\mu^i$
($i = 1, 2$) and a vector boson $A_\mu$; the hypermultiplet contains four real scalars $q$  and hyperinos $\zeta $;
the $N = 2$ vector multiplet has a vector field, two
spin-1/2 fermions and one real scalar field. The fermions of each of these multiplets
transform as doublets under the $SU(2)_R$ $R$-symmetry group of the $N = 2$
superalgebra. 
The ungauged theory is  determined in terms of real symmetric tensor $C_{IJK}$, $I,J,K=0,...,n_V$.
The vector multiplet scalars $h^I$ satisfy $C_{IJK}h^I h^J h^K=1$ which
define an $n_V$ dimensional hypersurface of scalars $\phi^x$ called a `very special real' manifold.

We consider the supersymmetry variation of fermions after turning on constant background values for vector fields and scalars.
Like in the four-dimensional case, the vector fields must be constant in order to to satisfy the  requirement $F_{\mu\nu}(A)=0$ coming from integrability of the 
vanishing gravitino transformation.
The supersymmetry transformations for the gravitino $\psi_\mu^i$, gauginos $\lambda^{xi}$ and hyperinos $\zeta^A$  take the following form
(see e.g. \cite{Gunaydin:1984ak,Bergshoeff:2004kh})
\bea
&& \delta \psi_\mu^i = \nabla_\mu\xi^i -g \kappa^2  A_\mu^I P_I^{ij}\xi_j  -{ig\over \kappa\sqrt{6}} \ \Gamma_\mu P^{ij} \xi_j\ ,
\nonumber\\
&& \delta \lambda^{xi}=-{ig\over 2}\Gamma^\mu A^I_\mu K^x_I \xi^i -{g\over \kappa^2}  P^{x ij}\xi_j+{g\over \kappa^2} W^x \xi^i\ ,
\label{mar}
\\
&& \delta \zeta^A={ig\over 2}\Gamma^\mu A^I_\mu k^X_I f^{iA}_X \xi_i +{g\over \kappa^2} {\cal N}^A_i \xi^i\ .
\nonumber
\eea
 $i=1,2$, $\xi_j=\epsilon_{ij} \xi^j$. 
 One can switch between $SU(2)$ and vector indices by using the relation
 \be
 A_i^{\ j}\equiv i \vec A\cdot \vec\sigma^{\ j}\ ,
 \ee
 where $ \vec\sigma $ are Pauli matrices.
The spinors obey the pseudo Majorana condition $\bar \xi^i =(\xi_i)^*\Gamma_0=\xi^{iT}C$.
$P^{ij}$ and $W$ depend on the constant backgrounds for the scalars $q^X$ and $\phi^x$.
For further notation and details  we refer to \cite{Gunaydin:1984ak,Bergshoeff:2004kh}.

In the Euclidean theory, we treat $\xi^i $ and $\bar \xi^i$ as independent spinors. 
We will define $\bar \xi^j= - i\xi_j$. 
The vanishing of the gravitino transformation 
then implies the two separate conditions
\bea
&& 0= \nabla_\mu\xi^i -i g \kappa^2  A_\mu^I P_I^{ij}\bar \xi^j  +m \ \Gamma_\mu P^{ij} \bar \xi^j\ ,
\nonumber\\
&& 0= \nabla_\mu\bar \xi^i +i g \kappa^2  \bar A_\mu^I \bar P_I^{ij}\xi^j  + m \ \Gamma_\mu \bar P^{ij} \xi^j\ ,
\label{mark}
\eea
with
\be
m\equiv {g\over \kappa\sqrt{6}} \ .
\ee
It should be noted that in particular cases the present Killing spinor equation simplifies. For example, in  gauged supergravity 
coupled to only vector multiplets described in  \cite{Gunaydin:1984ak} one has  $P^{ij}\propto \delta^{ij} $. 

Let us first look for spaces with maximal supersymmetry and consider solutions with $A^I_\mu=0$. The gaugino and hyperino equations give constraints on the values of the constant scalar fields.
Supersymmetric solutions with no restrictions on $\xi^i, \bar\xi^i$ require that
\be 
  P^{x ij}= W^x = {\cal N}^A_i =0\ .
\label{garaa}
\ee
These conditions are very similar to the ones appearing in studies of supersymmetric renormalization flows in AdS solutions of $N=2$ 5d gauged supergravity \cite{Ceresole:2001wi}.
The solution, which in particular depends on the specific scalar manifolds, 
represents fixed points of the renormalization group where the scalars are frozen.
For the  purpose of this work, one just needs to bear in mind that the possible constant values of the scalar fields will be given by the solutions to (\ref{garaa}),
which is to be found explicitly once the model is specified.

The scalar potential then simplifies to
\be
V=-\frac{4g^2}{\kappa^4} \, \vec P\cdot\vec {\bar P} 
\ee
where we assumed   $[P,\bar P]=0$.  
{} Let us now consider the gravitino equations. 
 The  integrability conditions give
\bea
&& 0 =  \big[\nabla_\nu,\nabla_\mu\big]\xi^i 
  +\kappa^2\, \frac{V}{12} \Gamma_{\mu\nu}  \   \xi^i \ ,
\nonumber\\
&& 0 =  \big[\nabla_\nu,\nabla_\mu\big]\bar \xi^i  
  +\kappa^2\, \frac{V}{12}  \Gamma_{\mu\nu}  \  \bar \xi^i \ .
\eea
By a similar calculation as in the previous section, we  get the conditions
\be
W_{\mu\nu\rho\sigma }=0\ ,\qquad R_{\mu\nu}=\kappa^2\,  \frac{2V}{3} g_{\mu\nu}\ ,
\ee
i.e. the space is Einstein-Weyl.
This implies that the space is locally isometric to $\mathbb{H}_5$ or $\mathbb{S}^5$ according to the case,
$V<0$ or $V>0$, where $V$ is to be evaluated at the scalar background. 
The space $\mathbb{S}^5$ has been used to carry out exact calculations of the partition function in   $N=1$ Super Yang-Mills theory  and superconformal theories  \cite{hosomichi2,Jafferis3}.

\medskip

Spaces of the form $X_4\times \mathbb{S}^1$  arise   by turning on  constant Wilson lines.
Define $t^{ij}=\kappa^2  A_5^I P_I^{ij}$, $\bar t^{ij}=\kappa^2  \bar A_5^I \bar P_I^{ij}$.
Similarly to the four-dimensional case,  we demand
\be
\p_5 \xi^i =\p_5 \bar \xi^i=0\ ,
\ee
so that
\be
ig  t^{ij} \bar \xi^j =  m  \Gamma_5 P^{ij}\bar \xi^j\ ,\qquad ig \bar   t^{ij}   \xi^j =  -m \Gamma_5\bar P^{ij}   \xi^i\ .
\label{hara}
\ee
The presence of $\Gamma_5$ reduces the number of supersymmetries by a factor 1/2.
To have no further restrictions on the spinors, we impose the following algebraic equations for the background values of scalars
and Wilson lines:
\be
ig  t^{ij}=  m   P^{ij} \ ,\qquad  ig \bar   t^{ij}  = -m \bar P^{ij} \ .
\ee
The remaining equations are
\bea
&&\nabla_{\hat\mu}\xi^i =-m \Gamma_{\hat\mu} P^{ij}\bar \xi^i \ ,
\nonumber\\
&& \nabla_{\hat\mu} \bar\xi ^i =-m \Gamma_{\hat\mu}\bar P^{ij} \xi^i \ .
\eea
The integrability condition then gives
\be
W_{\hat\mu \hat\nu \hat\rho\hat \sigma}= 0 \ ,\qquad R_{\hat\mu\hat\nu }=  6 m^2 P_k^{\, j}\bar P_{j}^{\, k} g_{\hat\mu\hat\nu }
= \frac{\kappa^2}{2}\ V\ g_{\hat\mu\hat\nu }\ ,
\ee
i.e. we get a space locally isometric to $\mathbb{S}^4\times \mathbb{S}^1$ or $\mathbb{H}^4\times \mathbb{S}^1$ according to the sign
of $V$.

\bigskip

It is interesting to compare with  \cite{tera}, where supersymmetric gauge theories on $\mathbb{S}^4\times \mathbb{S}^1$  are discussed.
In this work, maintaining supersymmetry required introducing by hand a new contribution in the Killing spinor equation containing a 
symmetric tensor $t^{ij}$. However, the construction of a supersymmetric  Yang-Mills action turns out to be problematic.
Despite some similarities, in the present approach the Killing spinor equation  has a  structure which is different from
the one proposed in \cite{tera}.
Our approach explains the origin of the tensor $t^{ij}$ and justifies this term in the gravitino transformation laws. 
Moreover, since the supergravity action coupled to vector multiplets
is,  by construction, supersymmetric, we expect that our approach also prescribes how to determine the 
complete supersymmetric and gauge invariant Yang-Mills action. We leave this interesting problem for future research.

\section{Supersymmetric spaces in six dimensions}

According to the group theoretic analysis of section 2, six dimensions 
is the highest dimension that can be considered 
for a consistent quantum theory with global supersymmetry based on simple supergroups.
Consequently, it is the highest dimension allowed from AdS/CFT correspondence,
an  example being the (2,0) superconformal field theory describing
the low energy dynamics of M5 branes.

\subsection{$F(4)$ gauged Supergravity} 

In order to find possible supersymmetric spaces in six dimensions, 
we will  employ Romans $F(4)$ gauged supergravity
 \cite{romans}. The theory includes an $SU(2)$ connection $A_\mu^{ij}$, an abelian connection $a_\mu$ 
and an antisymmetric form $B_{\mu\nu}$, with field strengths $F_{\mu\nu}(A)$, $f_{\mu\nu}(a)$ and $G_{\mu\nu\rho}$, respectively.
We consider configurations with $F_{\mu\nu}(A)=G_{\mu\nu\rho}=f_{\mu\nu}=0$ and we will allow only for a non-vanishing
flat $SU(2)$ field $A_{\mu}^{i j}$. In Lorentzian signature, fermions in the theory are symplectic Majorana. 
However, in Euclidean signature 
the symplectic Majorana condition is relaxed.   
The gravitino transformation law may be written as 
\be
\delta \psi_{\mu}^i = \nabla_\mu \xi^i  -ig A_{\mu \ j}^{\ i }\xi^j
+  T\ \Gamma_\mu \Gamma_7\xi^i \ ,
 \label{f6}
\ee
where
\be
T=-{1\over 8\sqrt{2}} (g \, e^{\phi/\sqrt{2}}+m\, e^{-3\phi/\sqrt{2}}) \ .
\ee
Note that the scalar  $\phi$ should be constant as follows from the vanishing of supersymmetric 
shifts of the four spin-$\frac{1}{2}$ fields $\chi_i$ in the gravity multiplet 
\be
\delta  \chi_i =\frac{1}{\sqrt{2}}\Gamma^\mu\partial_\mu \phi\xi_i+\frac{1}{4\sqrt{2}}\left(g \, e^{\phi/\sqrt{2}}-3 m
\, e^{-3\phi/\sqrt{2}}\right)\Gamma_7\,\xi_i\ .
\ee
Then $\delta \chi_i=0$ gives that 
\be
\phi= \frac{\sqrt{2}}{4}\ln \left(\frac{3m}{g}\right)\, , ~~ T=-\frac{1}{6\sqrt{2}} g\left(\frac{3m}{g}\right)^{1/4}\ .
\ee

\noindent
There are now two cases:
\\
{\bf {I. $\mathbf{A_{\mu\  j}^{\ i }=0}$}}
\\
The  integrability condition of (\ref{f6}) is written here as
\be
0 =  \big[\nabla_\nu,\nabla_\mu\big]\xi^i -T \bar T\, \Gamma_{\mu\nu}  \   \xi^i \ ,
\ee
and leads to 
\be
W_{\mu\nu\rho\sigma}=0\, , ~~~~R_{\mu\nu}=-3 T\bar T \, g_{\mu\nu} \ .
\label{wr6}
\ee
The only solution to (\ref{wr6}) is the round $\mathbb{S}^6$ (in the compact case, $T\bar T<0$) or $\mathbb{H}^6$ (in
the non-compact case, $T\bar T>0$). For  $\mathbb{S}^6$, the background enjoys  an $SO(7)$ isometry and, for $\mathbb{H}^6$, 
 $SO(6,1)$. The $SO(6,1)$ can be the bosonic part of a supersymmetry algebra whereas, 
 according to Nahm's classification, $SO(7)$ cannot. 
 


In Lorentzian space, the $F(4)$ gauged supergravity theory has anti-de Sitter solutions with $SO(5,2)\times SU(2)$
 bosonic symmetry, representing a subgroup of $F(4)$. This is of course in Nahm's list. 
 Nahm's classification also includes $SO(7)\times SO(2,1)$ -- case {\bf XII} in section 2 -- which is another real form for the bosonic subgroup of $F(4)$. \footnote{The other real forms of $F(4)$ are $SU(1,1)\times SO(7)$, $SU(2)\times SO(6,1)$ and $SU(2)\times SO(4,3)$ \cite{sorba2}. We thank
 Paul Sorba for clarification on this point.} 
Interpreting this $SO(7)$ as the symmetry of $\mathbb{S}^6$ implies that $SO(2,1)=SU(1,1)$ must arise as an internal R-symmetry.
This group is non-compact and would inevitably lead to ghosts.

In \cite{honda}, the authors claim to have constructed supersymmetric
Yang-Mills theories in $\mathbb{S}^d$ with
$d\leq 7$. This claim includes the cases of $\mathbb{S}^6$ and $\mathbb{S}^7$.
These spaces have $SO(7)$ and $SO(8)$ bosonic isometries, respectively.
According to Nahm's classification reviewed in section 2,
these symmetries are only present in the cases {\bf X}, ${\cal G}= osp(N|2)$, with
$N=8$ or ${\bf XII}$.
There exists no other superalgebra which
contains $SO(7)$ or  $SO(8)$ as bosonic isometries.
However, in both cases, the full bosonic symmetry also contains the
non-compact group $SO(2,1)$. The kinetic terms will
have to be invariant under this symmetry, which implies that the
theory necessarily contains ghosts.
In the notation of \cite{honda}, this class of theories seem to correspond to the cases 
called ``Class 2", and in {\it Euclidean} space the $SO(2,1)$ group should arise from
the generators ${\bar R}_{pq}$ with $p,q=7,8,9$.
This symmetry would lead to kinetic terms in the action with wrong signs. Unfortunately, the R-symmetry algebra is not derived in \cite{honda} and there is no discussion on which changes should be applied in going from Minkowski to Euclidean space. It would be interesting to clarify the structure of the superalgebras  for $\mathbb{S}^6$ and $\mathbb{S}^7$ in \cite{honda} 
and to see if they indeed correspond to cases {\bf X}, {\bf XII }   in Nahm's classification.


\medskip

In conclusion,  $\mathbb{S}^d$ with $d\geq 6$ 
cannot support supersymmetry.   We shall expand on the problems of $\mathbb{S}^d$ with $d\geq 6$  in section 5.2 and section 6.

\medskip

\vskip.2in
\noindent
{\bf {II. $\mathbf{A_{\mu\  j}^{\ i }\neq 0}$}}
\\ 
In this case, we may consider Wilson lines for an Abelian subgroup of the R-symmetry group $SU(2)$. For example, 
in the simplest
case, we may switch on a a flat  $A=A_{6\  j}^{\ i }dx^6=\frac{1}{2}A_{6}(\sigma_3)_j^idx^6$ $U(1)$ field. 
Then, the vanishing of the gravitino shifts
can be written as 
\bea
&&0= \nabla_6 \xi^i  -ig A_{6}(\sigma_3)_j^i\xi^j+  T\ \Gamma_6 \Gamma_7\xi^i \ ,
\nonumber\\
&&0= \nabla_{\hat \mu} \xi^i  +  T\ \Gamma_{\hat \mu} \Gamma_7\xi^i \, , ~~~\hat \mu =1,...,5\ .  \label{s1s}
\eea
These equations are solved then by $\xi^i=\xi^i(x^{\hat \mu })$ on $\mathbb{S}^1\times \mathbb{S}^5$. 
The corresponding superalgebra is $su(4|2)$, described by case {\bf II} of section 2.

\medskip

\subsection{Supersymmetric algebras in 6d}

One would like then to know why $\mathbb{S}^6$ fails to admit  supersymmetry.
In order to see this, let us recall that the full symmetry group of $\mathbb{S}^6$  is expected to be
$SO(7)\times R$ ($SO(7)$ from its isometry group and $R$ an R-symmetry group), which represents the bosonic (even) ${\cal{G}}_0$ subgroup of supersymmetry.
Then,  there should also exist an 
odd part ${\cal{G}}_1$, 
transforming in the spinorial representation of the even (bosonic) part of the supersymmetry algebra. 
To find ${\cal{G}}_1$, we recall
that 
\be
\{ {\cal{G}}_1,{\cal{G}}_1\} \subset {\cal{G}}_0\ .
\ee
As we are looking for supersymmetry, the odd generators should be fermionic and  so in 
 the spinorial representation of the 
even  $SO(7)\times R$ algebra. To make things simpler, we will consider first the case where the 
even part is just $SO(7)$.
In this case,  and since 
\bea
\bf 8\times 8=\bf 1_s+\bf 21_a+\bf 7_a+\bf 35_s\ ,
\eea
the fermionic anticommutator should close into  
 the $\bf 21$ 
(i.e., the generators $M_{mn}$ of the $SO(7)$),
\be
\{Q_\alpha,Q_\beta\}= \kappa({\sigma^{mn}}{\cal{C}}^{-1})_{\alpha\beta}M_{mn}\ ,
\ee
where  $(m,n,...=1,...,7), ~~~(\alpha,\beta,...=1,...8)$, 
$\kappa$ is an appropriate constant and $\sigma^{mn}=\frac{1}{4}[\gamma^m,\gamma^n]$
is the $SO(7)$ spinorial representation\footnote{
Standard Poincar\'e supersymmetry corresponds to the closure of the fermionic anticommutator in the $\bf 7$ 
representation of $SO(7)$, i.e.,
\be
\{Q_\alpha,Q_\beta\}=({\gamma^m}{\cal{C}}^{-1})_{\alpha\beta}P_m\, .
\ee }.
The generators $M_{mn}$ satisfy the $SO(7)$ algebra 
\be
[M_{mn},M_{kl}]=-\delta_{mk}M_{nl}+\delta_{ml}M_{nk}-
\delta_{nl}M_{mk}+\delta_{nk}M_{ml}\ .
\ee
Since $Q_\alpha$ are fermions, they transform under the spinorial representation as 
\be
[Q_\alpha,M_{mn}]=\frac{1}{2}\, {(\sigma_{mn})_\alpha}^\beta Q_\beta\ .
\ee
Now, all commutation relations have been defined and 
what remains to be checked is the Jacobi identity.  For the triplet $(M_{mn},Q_\alpha,Q_\beta)$ it reads
\be
[\{Q_\alpha,Q_\beta\},M_{mn}]+\{[M_{mn},Q_\alpha],Q_{\beta}\}+\{[M_{mn},Q_\beta],Q_{\alpha}\}=0\ ,
\ee
and  leads to  the conditions
\bea
0\!&\!=\!&\! \kappa \big{(}\sigma_{kl}{\cal{C}}^{-1}\big{)}_{\alpha\beta}\Big{(}\delta_{mk}M_{nl}-\delta_{ml}M_{nk}+
\delta_{nl}M_{mk}-\delta_{nk}M_{ml}\Big{)}\nonumber \\
&-& {1\over 2} 
\big{(}\sigma_{mn}\big{)}_{\beta\gamma}\big{(}\sigma^{kl}{\cal{C}}^{-1}\big{)}_{\gamma\alpha}
M_{kl}- {1\over 2} 
\big{(}\sigma_{mn}\big{)}_{\alpha\gamma}\big{(}\sigma^{kl}{\cal{C}}^{-1}\big{)}_{\gamma\beta}
M_{kl} \ .
\label{jj}
\eea
Obviously, this relation is the same for any group $SO(d)$ with odd part in the spinorial representation.
It is also clear that such a relation cannot be satisfied in general and it can only be valid  accidentally. 
This is indeed the case for the $SO(7)$ group. By using the following representation of the $SO(7)$ $8\times 8$
$\gamma$-matrices
\be
(\gamma_m)_{ab}=i\psi_{mab}\, , ~~~(\gamma_m)_{8a}=i \delta_{ma}
\ee
where $\psi_{mab}$ are the octonionic structure constants \cite{fk} and the relation
\be
\psi^{abc} \psi_{dhc}=\delta^a_d\delta^b_h-\delta^b_d\delta^a_h-\frac{1}{3!}{\epsilon^{ab}}_{dgijk}\psi^{ijk}\, ,
\ee
we find that miraculously (\ref{jj}) is satisfied. However, there are also other Jacobi identities which should be 
satisfied. Among these, it is straightforward to check that 
\be
[\{Q_\alpha,Q_\beta\},Q_\gamma]+[\{Q_\alpha,Q_\gamma\},Q_{\beta}]+[\{Q_\beta,Q_\gamma\},Q_{\alpha}]=0\ ,
\label{jacob2}
\ee
fails to be satisfied.  The Jacobi identity can be satisfied if $SO(7)$ is extended in an appropriate way. 
In particular, the appropriate extension turns out to be $SO(7)\times SU(1,1)$ 
and in this case it has be  proven that a superalgebra exists \cite{kac,rn,wit-van}. It is defined by the commutation relations
\begin{align}
&[T_i,T_j]=ic_{ij} ^k T_k\, , \qquad\qquad\, \,  [T_i,M_{mn}]=0\ ,
\\
&[M_{mn},M_{kl}]=-\delta_{mk}M_{nl}+\delta_{ml}M_{nk}-
\delta_{nl}M_{mk}+\delta_{nk}M_{ml}\, ,\\
&[T_i,Q^a_\alpha]=\frac{1}{2}{(\tau_i)^a}_b Q_\alpha^b\, , \qquad [M_{mn},Q^a_\alpha]={(\sigma_{mn})_\alpha}^\beta
Q^a_\beta\ ,\\
&\{Q^a_\alpha,Q^b_\beta\}=2 C^{(8)}_{\alpha\beta}\big(C^{(2)}\tau^i\big{)}^{ab}T_i
+\frac{2}{3}{C^{(2)}}^{ab}\big{(}C^{(8)}\sigma^{mn}\big{)}M_{mn} \ ,
\end{align}
where $(i=1,2,3)$ , $(\tau^i)$ are the fundamental representation of $SU(1,1)=SO(2,1)$ and $C^{(2)}\, (=i\tau^2),\, C^{(8)}$ are the
$2\times 2$ and $8\times 8$ charge conjugation matrices. Of course, we recognize here the exceptional 
 $F(4)$ superalgebra. Note that the odd generators of the algebra are in the $(\bf{8},\bf{2})$ representation 
of the even $SO(7)\times SU(1,1)$ and therefore it is a supersymmetry algebra. This is case  {\bf XII} in section 2.
There is another extension by which the Jacobi identity can be satisfied, namely the $osp(7|2)$ superalgebra, also
with bosonic group $SO(7)\times SU(1,1)$, corresponding to case {\bf X}.
However, in this case the odd generators are not in the spinorial representation of the $SO(7)$ isometry group.
For the $F(4)$ superalgebra, the odd generators are in the spinorial representation of $SO(7)$ but
one still has the problem  that the R-symmetry
group is the non-compact $SU(1,1)$, which has indefinite metric (Cartan-Killing metric of signature $1$) and therefore 
any theory invariant under $F(4)$ supersymmetry will necessarily have ghosts. This is the reason why the $F(4)$ of case {\bf XII}
cannot be used as a possible superalgebra in a ghost-free supersymmetric theory on  $\mathbb{S}^6$. 

One may also ask if there are {\it superconformal} theories on $\mathbb{S}^6$. If that was the case, there should be a 
superalgebra with even part 
containing $SO(7,1)$. However, a simple inspection of the classification of possible superalgebras reveals that this is not possible. Therefore,
$\mathbb{S}^6$ does not admit superconformal theories and of course, this is also the case for Euclidean $\mathbb{R}^6$, which should share the same
$SO(7,1)$ symmetry.

It is instructive to compare with the lower dimensional cases where we have supersymmetry. For example, 
let us now explicitly demonstrate  why  $\mathbb{S}^5$ does admit supersymmetry.
The isometry group of $\mathbb{S}^5$  is $SO(6)$ and we can consider the generators $T_\mu^\nu$ 
(in the $\bf 15$  of 
$SO(6)\simeq SU(4)$) to satisfy
\be
[T_m^n,T_k^l]=\delta_k^n T_m^l-\delta_m^l T_k^n\ ,
\ee
where $m,n,k,l=1,2,3,4$.
We can take the odd part of the superalgebra of which $SO(6)$ is the even part to be generated by $Q_m$ and $Q^n$ 
in the $\bf 4$ and $\bf \bar 4$ spinorial representations of $SO(6)$,
\be
[T_m^n,Q_k]=\delta_k^n Q_m\, , ~~~[T_m^n,Q^k]=-\delta^k_m Q^n\ .
\ee
Then, since 
\be
\bf 4\times \bf 4=\bf 6 +\bf 10\, , ~~~~\bf 4 \times \bf \bar 4 =\bf 1 + \bf 15\ ,
\ee
we see that necessarily 
\be
\{Q^m,Q^n\}=0\, , ~~~\{Q_m,Q_n\}=0\, , ~~~\{Q^m,Q_n\}=\beta T_n^m+\delta^m_n Z\ ,
\ee
where $Z$ is an $SO(6)$ singlet and $\beta$ a constant, which is specified from Jacobi identity to be $\beta=0$.
Then the operators $T^m_n, Z,Q_m,Q^n$  generate the supergroup $su(4|1)$ and, since the odd generators are in the 
spinorial representation of the even $SO(6)\times U(1)$ algebra, it is a supersymmetric algebra.
Therefore, we see that the existence of supersymmetry on $\mathbb{S}^5$ is  basically due to the Lie algebra isomorphism
$SO(6)\simeq SU(4)$, which permits that $SO(6)$ also arises in the  $su(n|N)$ series of supergroups.

\section{Supersymmetry in $d>5$}

According to the discussion in section 2 and above, it is not possible to have supersymmetry on $d$-spheres with $d>5$. 
Field theories with rigid supersymmetry exist up to $d=10$.
One important case that can be ruled out immediately is supersymmetry on $S^{10}$, due to the fact that the minimal spinor
representation for $SO(11)$ is 32. In particular, $N=1$ 10d super Yang-Mills theory cannot be put on $S^{10}$.

More generally, in $d>5$, one of the problems is that, as explained, the  Jacobi conditions  (\ref{jj}), (\ref{jacob2}) fail to be satisfied 
for odd generators in the  spinorial representation of the even  
($SO(d)$ (or $SO(d+1,1)$) part, unless the bosonic symmetry is extended in an appropriate way.
To identify the odd part of the superalgebra, let us assume that the corresponding
generators $Q_\alpha$ transform in a particular representation $\Delta$ as 
\bea
&&\{Q_\alpha,Q_\beta\}= \kappa\big{(}\Delta^{\mu\nu}\big{)}_{\alpha\beta}M_{\mu\nu}\ ,\\
&&
[Q_\alpha,M_{\mu\nu}]=\frac{1}{2}\, {(\Delta_{\mu\nu})_\alpha}^\beta Q_\beta\ .
\eea
Then, the Jacobi identity (\ref{jj}) gives
\bea
0\!&\!=\!&\!\kappa\big{(}\Delta_{\kappa\lambda}\big{)}_{\alpha\beta}\Big{(}\delta_{\mu\kappa}M_{\nu\lambda}-\delta_{\mu\lambda}M_{\nu\kappa}+
\delta_{\nu\lambda}M_{\mu\kappa}-\delta_{\nu\kappa}M_{\mu\lambda}\Big{)}\nonumber \\
&-& 2
\big{(}\Delta_{\mu\nu}\big{)}_{\beta\gamma}\big{(}\Delta^{\kappa\lambda}\big{)}_{\gamma\alpha}
M_{\kappa\lambda}- 
\big{(}\Delta_{\mu\nu}\big{)}_{\alpha\gamma}\big{(}\Delta^{\kappa\lambda}\big{)}_{\gamma\beta}
M_{\kappa\lambda} \ .
\label{jjj}
\eea
The solution to this condition specifies the allowed representation for the odd generators. 
It is easy to check that 
(\ref{jjj}) is solved for 
\be
\big{(}\Delta_{\mu\nu}\big{)}_{\alpha\beta}=\delta_{\mu\alpha}\delta_{\nu\beta}-
\delta_{\mu\beta}\delta_{\nu\alpha}\ ,
\ee
and $\kappa=1$. Thus, $(M_{\mu\nu},Q_\alpha)$ form a superalgebra if $Q_\alpha$  transforms 
in the vectorial representation of the even $SO(d)$ part. This superalgebra is the $osp(d|2)$.
This appears as  case {\bf X} in  section 2, except for the crucial difference that here $SO(d)$ appears as
the isometry and $SO(2,1)$ as the R-symmetry, whereas in section 2 is the opposite.
Because  for $d>5$   the odd generators are not in 
the spinorial representation of the isometry group,  $osp(d|2)$ is not a supersymmetry algebra  (and, similarly, $osp(d\!+\!1,\!1|2)$ is not 
a superconformal algebra). 
Moreover, the R-symmetry is represented by the non-compact $SO(2,1)$ group, therefore invariant Lagrangians will contain 
ghosts.
We emphasize the double role that the even part $SO(d)\times SO(2,1)$ of $osp(d|2)$ can play: either $SO(2,1)$ is the isometry and $SO(d)$ the R-symmetry,
or the opposite, $SO(d)$ is the isometry and $SO(2,1)$ the R-symmetry. Clearly, only the first case makes sense ({\bf X} of section 2)
 if one wishes to have a ghost-free theory.

In general, for any $m,n$, the even part of the $osp(m|n)$ is $SO(m)\times Sp(n)$ and the odd part 
is in the $(m,n)$ representation of $SO(m)\times Sp(n)$. Therefore it might  seem that these difficulties would also 
apply to lower dimensions. However, this is not true. 
In fact, 
(\ref{jj}) holds for certain specific cases. 
In particular, (\ref{jj}) `accidentally' holds  in $d\leq 5$ 
and the reason lies on the various Lie algebra isomorphisms  of the orthogonal groups
  with unitary and symplectic 
groups. Although orthogonal $SO(d)$ algebras are expected in the $osp(d|2)$ superalgebras which are not supersymmetries, some 
of them, due to isomorphisms, also appear in $su(d|N)$ superalgebras. These isomorphisms are
\bea \label{isom}
&SO(2)\simeq U(1)\, , ~~~SO(3)\simeq SU(2)\, , ~~~ SO(4) \simeq SU(2)\times SU(2)\, , & \nonumber\\
&  SO(5) \simeq USp(4)\, , ~~~ SO(6) \simeq SU(4) \ ,& \nonumber\\
&SO(5,1)\simeq SU^\ast(4)\, , ~~~ SO(4,1)\simeq USp(2,2)\, ,\ \ \ SO(3,1)\simeq SL(2,\mathbb{C}),\ \ \  SO(2,1) \simeq SU(1,1)\ ,&
\eea
and allow for the existence of supersymmetries on $d$-spheres with $d\leq 5$. The missing case $SO(6,1)$ 
is not due to some isomorphism but rather, as noticed already, 
 due to the fact that $SO(6,1)\times SU(2)$ is a real form of $F(4)$.


\medskip

In conclusion, Euclidean field theories with rigid supersymmetry cannot be consistently defined on
round spheres $\mathbb{S}^d$ if $d>5$.
In particular, in $d=6$ a superalgebra exists but the R-symmetry is non-compact leading to ghosts states.
Superconformal theories cannot be consistently defined on Euclidean $\mathbb{S}^d$ with $d>5$, nor on any space conformal
to $\mathbb{S}^d$ such as Euclidean $\mathbb{R}^d$ or $\mathbb{S}^{d-1}\times \mathbb{S}^1$ with $d>5$.

\vskip .3in
\noindent
{\bf Acknowledgment}: We would like to thank Jaume Gomis and Paul Townsend 
for useful discussions, Paul Sorba for correspondence and Gabriele Tartaglino-Mazzucchelli for remarks. 
We also thank the Perimeter Institute for Theoretical Physics for hospitality, where part of this work was carried out.
Research at Perimeter Institute
is supported by the Government of Canada through Industry Canada and by the Province of
Ontario through the Ministry of Economic Development and Innovation.
AK is supported by  the “ARISTEIA” Action of the “Operational Programme Education
and Lifelong Learning” and is co-funded by the European Social Fund (ESF) and National Resources.
JR acknowledges support from projects  FPA 2010-20807, 2009SGR502.


\end{document}